\documentclass{ws-procs975x65}
\def\bb {\begin {eqnarray}}
\def\ee {\end {eqnarray}}
\begin{document}

\title{Links and Quantum Entanglement
\footnote{Proceedings of the Conference in honour of Murray
Gell-Mann's 80th birthday (24-26 Feb 2010, NTU, Singapore),
pg.646-660.}}

\author{Allan I. Solomon}

\address{Department of Physics and Astronomy, Open University, \\
Milton Keynes, MK7 6AA, U.K.\\
E-mail: a.i.solomon@open.ac.uk\\
and\\
LPTMC, University of Paris VI, 75252, Paris, France
}
\author{Choon-Lin Ho}
\address{Department of Physics, Tamkang University, Tamsui
251, Taiwan, R.O.C.\\
and\\
Department of Physics, and Center for Quantum Technologies,
National University of Singapore, 117543 Singapore\\
E-mail: hcl@mail.tku.edu.tw }

\begin{abstract}
We discuss the analogy between topological entanglement and quantum entanglement, particularly  for tripartite quantum systems.
We illustrate our approach by first discussing two clearly (topologically) inequivalent systems of
three-ring links: The Borromean rings, in which the removal of any one link leaves
the remaining two non-linked (or, by analogy, non-entangled); and an inequivalent
system (which we call the NUS link) for which the removal of any one link leaves
the remaining  two linked (or, entangled in our analogy). We introduce unitary representations for the appropriate Braid Group ($B_3$) which produce the related quantum entangled systems. We finally remark that these two quantum systems, which clearly possess inequivalent entanglement properties, are locally unitarily equivalent.
\end{abstract}


\bodymatter
\def\figsubcap#1{\par\noindent\centering\footnotesize(#1)}
\begin{figure}[b]%
\begin{center}
  \parbox{2.1in}{\epsfig{figure=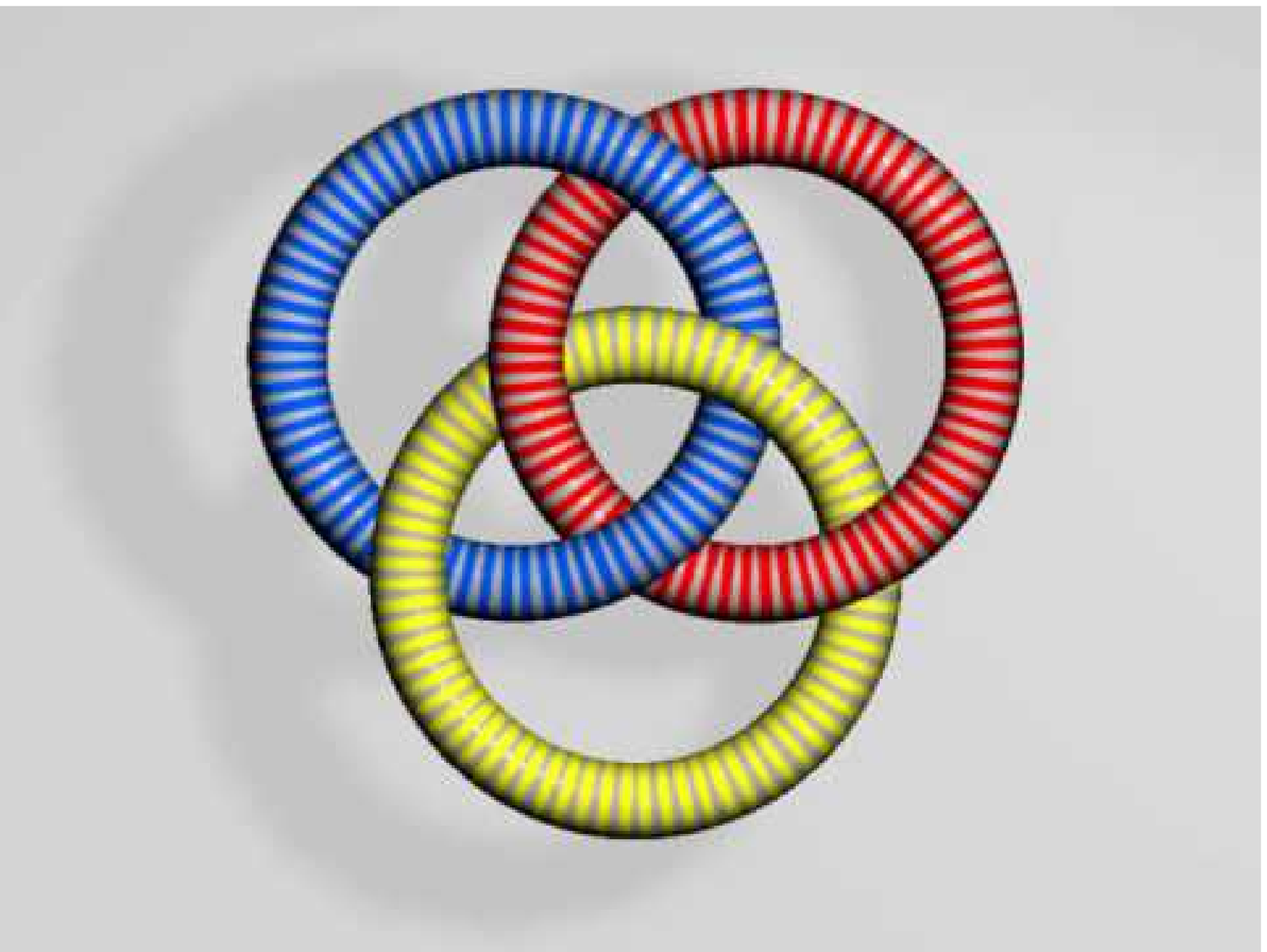,width=2in}\figsubcap{a}}
  \hspace*{4pt}
  \parbox{2.1in}{\epsfig{figure=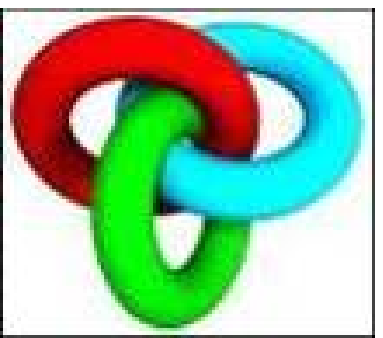,width=1.7 in}\figsubcap{b}}
  \caption{Two three-ring links. (a) Borromean Rings  (b) NUS Link}%
  \label{fig1.2}
\end{center}
\end{figure}
\section{Introduction:  The  Borromean Rings and the NUS Link}\label{intro}

In this note we shall explore the analogy between topological links and the
 quantum entanglement of tripartite systems.
In the figures  Fig.~\ref{fig1.2}(a) and \fref{fig1.2}(b), we give
examples of two different three-ring links.

The first, Fig.~\ref{fig1.2}(a), represents the celebrated
Borromean rings. This link has the property that removing any ring
leaves the remaining two rings unlinked (non-entangled). The
second,  \fref{fig1.2}(b) , which we call for brevity the NUS link
as it is part of the logo of the National University of Singapore,
has the converse property; removing any ring still leaves the two
remaining linked (entangled).

These two links recall the following tripartite quantum states:
  The Greenberger-Horne-Zeilinger (GHZ) state\cite{GHZ}, which
  is simply a tripartite extension of the bipartite Bell state
  $(1/\sqrt 2)(|0,0,\rangle+|1,1\rangle)$ ,
    \begin{equation}\label{GHZ}
|GHZ\rangle = (1/\sqrt 2)(|0,0,0\rangle+|1,1,1\rangle),
    \end{equation} and
 \begin{equation}\label{phi}
 |\phi \rangle =
 (1/2)(|0,0,0\rangle+|0,1,1\rangle+|1,0,1\rangle+|1,1,0\rangle).
 \end{equation}
 In the first case, measuring any subspace state as $|0\rangle$ ({\it resp.}
 $|1\rangle$) leads to the non-entangled state $|0,0\rangle$({\it resp.}
 $|1,1\rangle$); while in the second case a similar determination always
 leads to a (maximally) entangled  bipartite state (Bell state).

The mathematical representation of links is made via Braid Groups, introduced
by  Artin\cite{artin}.  To pursue the quantum entanglement analogy further, we
 first discuss braid groups, with an introductory reminder of a presentation of the
closely-related symmetric group.  Then, in order to apply these
ideas in quantum theory, we discuss their unitary representations,
which we take to act on the qubit spaces.
\section{Braid Groups and Links}
\subsection{Symmetric Group}
\begin{figure}[b]%
\begin{center}
  \parbox{2.1in}{\epsfig{figure=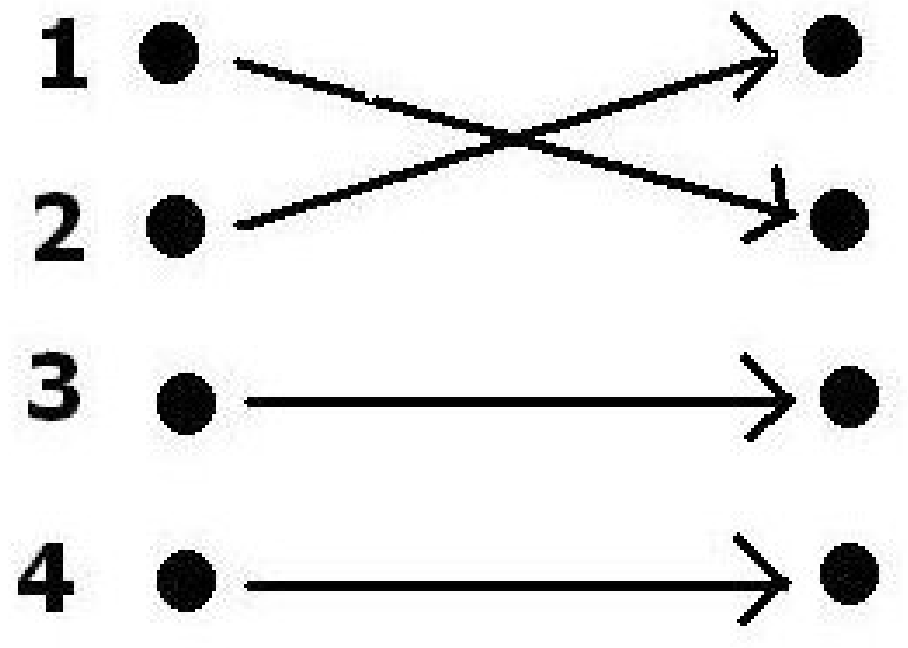,width=1.5 in}\figsubcap{a}}
  \hspace*{4pt}
  \parbox{2.1in}{\epsfig{figure=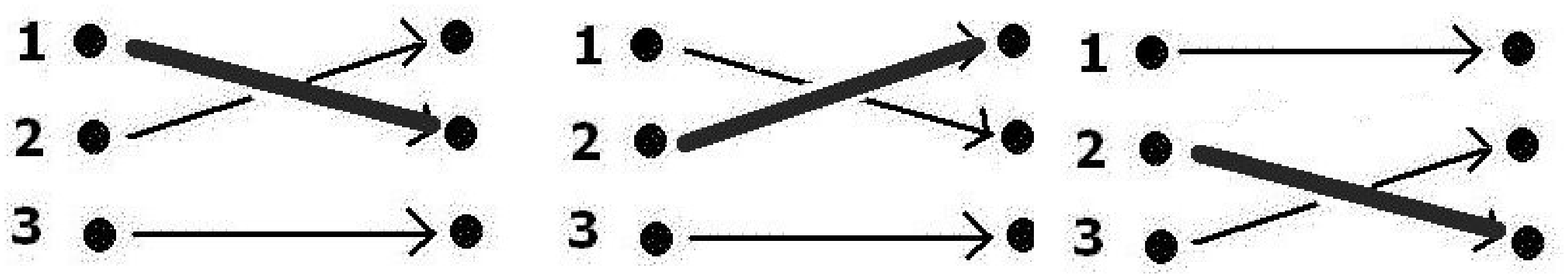,width=3 in}\figsubcap{b}}
  \caption{Elements of $S_4$ and $B_3$. (a) A transposition in $S_4$.
  (b) Elements $\sigma_1, \sigma_1^{-1}, \sigma_2$ of $B_3$.}%
  \label{s4b3}
\end{center}
\end{figure}

The symmetric group $S_n$ (sometimes called the permutation group)
is defined as the the set of $n!$ permutations on $n$ distinct objects,
combining according to the rule illustrated by
\begin{equation}\label{perm}
  \left(
    \begin{array}{c}
      1\;\;\;2\;\;\;3\;\;\;4 \\
       3\;\;\;1\;\;\;2\;\;\;4 \\
    \end{array}\right) \left(
    \begin{array}{c}
      1\;\;\;2\;\;\;3\;\;\;4 \\
       1\;\;\;3\;\;\;2\;\;\;4 \\
    \end{array}\right)
    =
     \left(\begin{array}{c}
      1\;\;\;2\;\;\;3\;\;\;4 \\
       2\;\;\;1\;\;\;3\;\;\;4 \\
    \end{array}\right)
    \end{equation}
for the case of $S_4$.  A diagrammatic representation of the resultant permutation
is found in Figure {\ref{s4b3}}(a).
The symmetric group $S_n$ has a presentation in terms of $n-1$ adjacent transpositions
\footnote{The right-hand side of Eq.(\ref{perm}) is an adjacent transposition.},
$\{s_i \; i=1\dots n-1\}$ where $s_i$ sends the $i$ to $i+1$ and $i+1$ to $i$.
This rather mysterious presentation is:
\bb
s_i s_j &=& s_j s_i \;\;\;\;\; |i-j|>1 \\
s_i s_i &=& I  \label{square}\\
s_i s_{i+1}s_i &=& s_{i+1} s_{i}s_{i+1} \label{yb}
\ee
where Eq.(\ref{yb}) plays an important role in the generalization to the {\em Braid group},
in which context it is known as the {\em braiding relation} or the {\em Yang-Baxter condition}.

\subsection{Braid group}
The braid group is like the symmetric group, but in three dimensions,
so one must imagine the arrows joining the elements of a permuted set of points to go ``over''
 or ``under'' each other.  Intuitively, each element of the braid group $B_n$ is one way of
 joining $n$ points to another $n$ points by strings. (For an expanded version of this
 intuitive definition see Reference{\cite{braid} .)

The braid group $B_n$ has a presentation in terms of $n-1$ generators $\sigma_i$.  This  (defining) presentation is:
\bb
\sigma_i \sigma_j &=& \sigma_j \sigma_i \;\;\;\;\; |i-j|>1  \label{1-2}\\
\sigma_i \sigma_{i+1}\sigma_i &=& \sigma_{i+1}
\sigma_{i}\sigma_{i+1} \label{yb1} \ee Note that the constraint
Eq.(\ref{square}) is absent; this absence leads to all the Braid groups being of infinite order. Eq.(\ref{yb1}) is known as the {\em
braiding relation} or the {\em Yang-Baxter} condition}. A
diagrammatic representation of the elements $\sigma_1$ and
$\sigma_1^{-1}$, as well as  the second generator $\sigma_2$,  of $B_3$ is
given in Figure \ref{s4b3}(b). This group  is the main example
that we discuss in this note, although for simplicity and
illustration  we start by discussing the group $B_2$, which has
only one generator, and no braiding condition to satisfy; it is
isomorphic to the infinite cyclic group, equivalently ${\cal Z}$,
the set of integers under addition.

\subsection{Knots and Links}
Of particular interest to us is the fact that, as shown by Alexander\cite{alexander},
{\em all} knots and links may be obtained from elements of a braid group by the simple
expedient of joining the the ``dots''; that is, join $1$ to $1$, $2$ to $2$, and so on.

\begin{figure}[t]
\begin{center}
\psfig{file=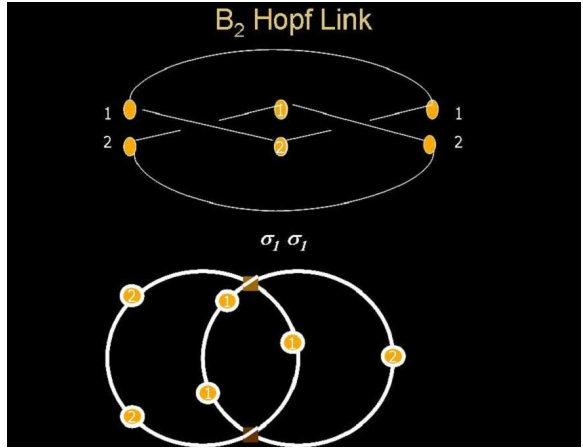,width=3 in}
\end{center}
\caption{In $B_2$, $\sigma_1^2$ produces the Hopf Link.}
\label{hopf}
\end{figure}

For the braid group $B_2$ with one generator $\sigma_1$, we can see that performing
the action using the element $\sigma_1^{2}$  gives the Hopf Link, as in Figure \ref{hopf}.

\begin{figure}[b]%
\begin{center}
  \parbox{2.1in}{\epsfig{figure=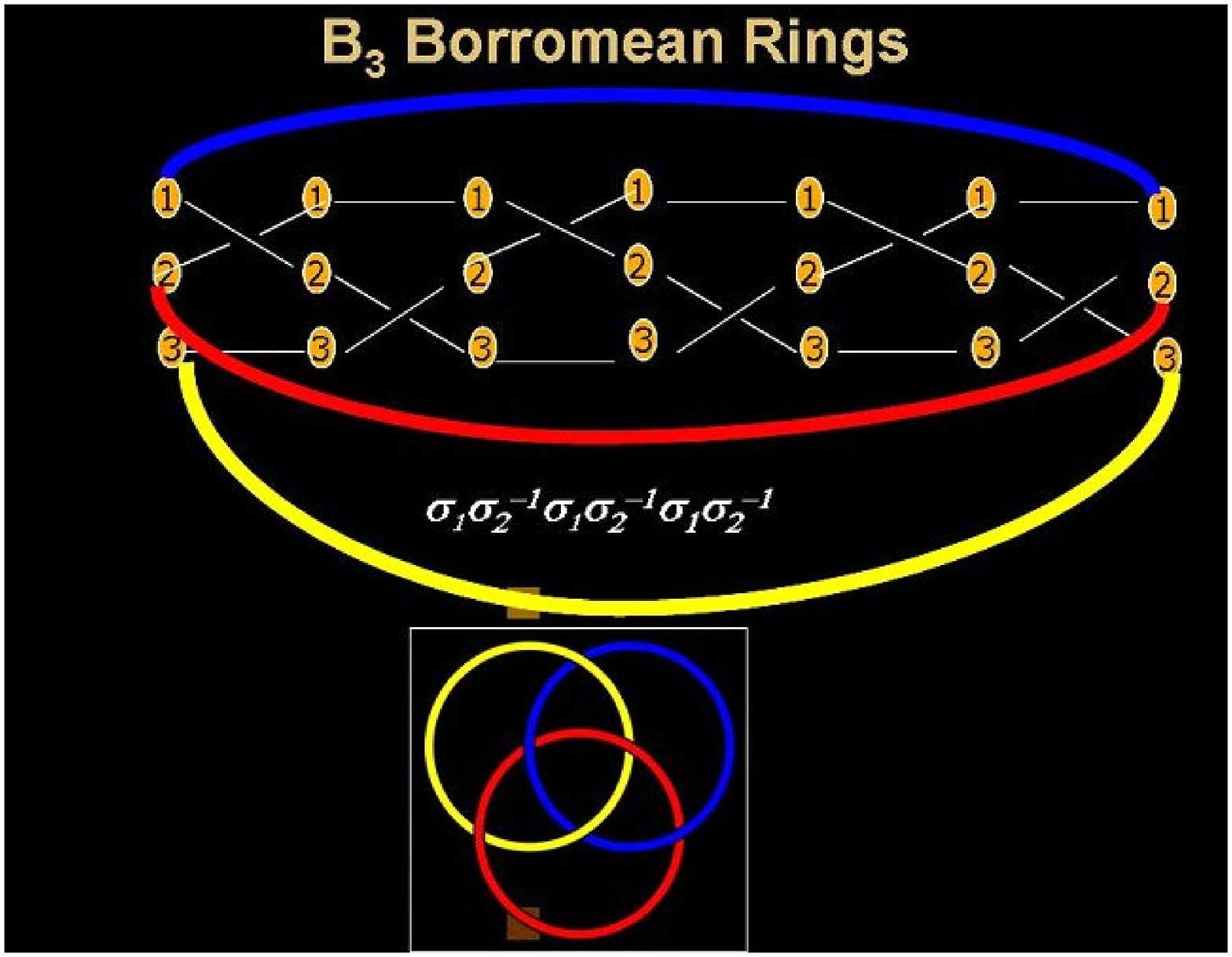,width=2.1 in}\figsubcap{a}}
  \hspace*{4pt}
  \parbox{2.1in}{\epsfig{figure=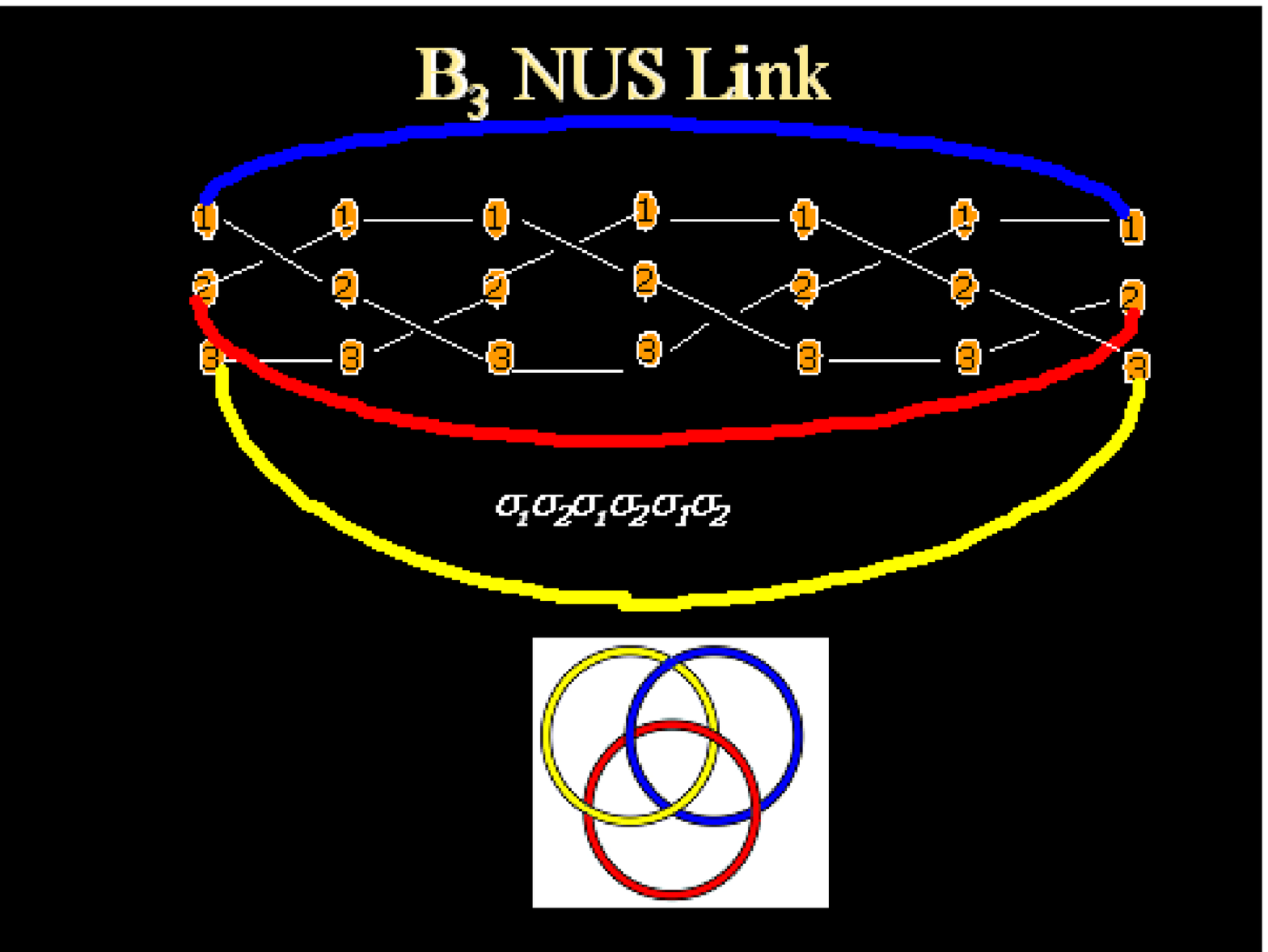,width=2.2 in}\figsubcap{b}}
  \caption{Two links from $B_3$. (a) Borromean Rings produced from an element of
  $B_3$. (b) NUS Link produced from an element of $B_3$.}%
  \label{3links}
\end{center}
\end{figure}

For the braid group $B_3$ with two generators $\sigma_1$ and $\sigma_2$, we can
see that performing this action with the braid element
\begin{equation}\label{b3borro}
    \sigma_1 \sigma_2^{-1}\sigma_1 \sigma_2^{-1}\sigma_1 \sigma_2^{-1}
\end{equation}
produces the Borromean rings, as in Figure \ref{3links}(a).

On the other hand,   the braid element
\begin{equation}\label{b3nus}
    \sigma_1 \sigma_2 \sigma_1 \sigma_2\sigma_1 \sigma_2
\end{equation}
corresponds to the NUS link, as in Figure \ref{3links}(b).
\section{Unitary Representations of braid groups and entanglement}
 In order to relate the action of the braid group to unitary transformations on
 quantum systems, we adopt the following procedure:
\begin{romanlist}[(iii)]\label{list}
\item we associate each initial point of the braid group with a qubit (e.g. for $B_3$
there are 3 initial points and therefore we may represent unitary
action on a three-qubit system);
\item for a braid word of the form $g^n$ we shall assume that the quantum entanglement
is generated by the unitary representative  $\hat{g}$;
\item to simulate the closure of the action of a braid word, say $g^n$, to form
a {\em link}, the unitary matrix ${\hat{g}}^n$ must equal $I$ (up to a phase factor).
\end{romanlist}
 A generic unitary representation of the braid group which satisfies the  relation
 Eq.(\ref{1-2})  can in principle be obtained  from the following:
 \begin{equation}\label{genrep}
\hat{\sigma}_i = I \times \cdots \times U \times I \cdots \times I
 \end{equation}
where
 $I=\left[
     \begin{array}{cc}
       1 & 0 \\
       0 & 1 \\
     \end{array}
   \right]$
   and $U$ is a $4 \times 4$ unitary matrix occupying the $(i,i+1)$ position in the product.
Of course it is more difficult to satisfy Eq.(\ref{yb1}), the
braiding, or Yang-Baxter, relation.  We describe representations
for $B_2$ and $B_3$ in the following.
\subsection{The Hopf link and entanglement}
   In a sense finding a unitary representation for $B_2$ is a trivial exercise, as in this
case there are effectively no relations on the single generator
$\sigma_1$.  Thus any unitary matrix will do. For our purpose we
require a $4 \times 4$ unitary matrix - since it is acting on the
two-qubit space.  We define a   unitary transformation  matrix as
follows\footnote{The multiplicative phase factor is necessary to
ensure a genuine representation of $B_2$, as in its absence the
representation would be {{\em non-faithful}}, and  finite
dimensional ($Z_2$).}:
   \begin{equation}\label{utheta}
  \hat{\sigma}_1\equiv \frac{e^{i \theta}}{\sqrt 2}\left[  \begin{array}{cccc}
    1 & 0 & 0 & 1 \\ 0 & 1 & 1 & 0 \\0 & 1 & -1 & 0 \\1 & 0 & 0 & -1 \\
    \end{array} \right]
    \; \; \; \; \; \; (\theta/\pi \; \; {\rm irrational}).
    \end{equation}
The braid word word corresponding to the Hopf link is ${\sigma_1}^2$ so following the
procedure as in \ref{list}(ii) outlined above,  our choice of  unitary representative
$\hat{\sigma}_1$  is the generator of entanglement, and produces a maximally entangled
(Bell) state from a (generic) non-entangled state,
\begin{equation}\label{b2bell}
\hat{\sigma_1}|0,0\rangle=\frac{\exp(i \theta)}{\sqrt{2}}(|0,0\rangle+|1,1\rangle).
\end{equation}
Note that $ {\hat{\sigma}_1}^2= e^{2i\theta}I$, satisfying condition \ref{list}(iii).

\subsection{Unitary representations for $B_3$}

\subsubsection{The NUS link and entanglement}
   Using the matrix $U$ of Reference \cite{ge} (where it is defined however without
   the phase factor) we define
   \begin{equation}
  U\equiv \frac{e^{i \theta}}{\sqrt 2}\left[
    \begin{array}{cccc}
    1 & 0 & 0 & -1 \\ 0 & 1 & -1 & 0 \\ 0 & 1 & 1 & 0 \\  1 & 0 & 0 & 1 \\
     \end{array}
   \right]
   \end{equation}
   where $\theta/\pi$  is irrational but otherwise arbitrary, as above. The representation
   for $B_3$ is
   \begin{equation}\label{b3rep}
  \hat{\sigma}_1=U\times I, \; \; \; \hat{\sigma}_2= I \times U.
   \end{equation}
   One may  verify that the braiding relation Eq.(\ref{yb1}) is satisfied.
   As in Eq.(\ref{b3nus}), the braid word $(\sigma_1 \sigma_2)^3$ produces  the NUS link.
   Following the recipe above, we note that $(\hat{\sigma_1} \hat{\sigma_2})^3$ is indeed
   the $8 \times 8$ unit matrix (up to a non-vanishing phase factor); and the generator of
   entanglement for this link $\hat{\sigma_1}\hat{ \sigma_2}$ produces the state
   $|\phi\rangle$ of Eq.(\ref{phi}) (up to the phase factor $e^{2 i \theta}$)
   \begin{equation}
  \hat{\sigma_1}\hat{ \sigma_2}|0,0,0,\rangle=\exp(2 i \theta)|\phi\rangle.
   \end{equation}

\subsubsection{Entanglement and the Borromean rings}
We use a different representation  for the Borromean rings in order to to obtain the  GHZ state directly.
Following  the procedure detailed in \cite{HSO} we use the Jones representation\footnote{In what follows we omit the explicit irrational phase factor needed to ensure the faithfulness of the representation.}
\begin{eqnarray}
\hat{\sigma}_i=Ah_i + A^{-1} I,\nonumber\\
\hat{\sigma_i}^{-1}=A^{-1}h_i + AI.\label{JonesRep}
\end{eqnarray}
We choose  $A=\exp(3\pi i/8)$, and the matrices $h_1$ and $h_2$ as follows:
\begin{eqnarray}
h_1
 =\sqrt{2} \left(
\begin{array}{cccccccc}
  1 & 0 & 0 & 0 & 0 & 0 & 0 & 0\\
  0 & 1 & 0 & 0 & 0 & 0 & 0 & 0\\
  0 & 0 & 1 & 0 & 0 & 0 & 0 & 0\\
  0 & 0 & 0 & 1 & 0 & 0 & 0 & 0\\
  0 & 0 & 0 & 0 & 0 & 0 & 0 & 0\\
  0 & 0 & 0 & 0 & 0 & 0 & 0 & 0\\
  0 & 0 & 0 & 0 & 0 & 0 & 0 & 0\\
  0 & 0 & 0 & 0 & 0 & 0 & 0 & 0
\end{array}\right).
\end{eqnarray}
and
\begin{eqnarray}
h_2
 =\frac{1}{\sqrt{2}} \left(
\begin{array}{cccccccc}
  1 & 0 & 0 & 0 & 0 & 0 & 0 &-1\\
  0 & 1 & 0 & 0 & 0 & 0 &-1 & 0\\
  0 & 0 & 1 & 0 & 0 &-1 & 0 & 0\\
  0 & 0 & 0 & 1 &-1 & 0 & 0 & 0\\
  0 & 0 & 0 &-1 & 1 & 0 & 0 & 0\\
  0 & 0 &-1 & 0 & 0 & 1 & 0 & 0\\
  0 &-1 & 0 & 0 & 0 & 0 & 1 & 0\\
  -1& 0 & 0 & 0 & 0 & 0 & 0 & 1
\end{array}\right).
\end{eqnarray}
Then it may be verified that $\hat{\sigma_1}$ and $\hat{\sigma_2}$ satisfy Eq.(\ref{yb1}).  The Borromean link is defined by the braid word given in Eq.(\ref{b3borro}), and additionally  the criterion of \ref{list}(iii) is satisfied, since
$
 (\hat{\sigma}_1\hat{\sigma}_2^{-1})^3
$
equals the identity up to a phase factor.

Applying the braid word entanglement generator, in this case $\hat{\sigma}_1\hat{\sigma}_2^{-1}$,  to the fiducial ground state $|0,0,0\rangle$, we obtain the  GHZ state
\begin{equation}
\hat{\sigma}_1\hat{\sigma}_2^{-1}|0,0,0\rangle =
\frac{1+i}{2}(|0,0,0\rangle+|1,1,1\rangle).
\end{equation}

\section{Conclusions: Local Unitary Equivalence}
This note has emphasized the analogy between topological entanglement in the form of links, and quantum entanglement.\footnote{This analogy has also been remarked upon  by, among others,  Kauffman and Lomonaco\cite{kl}, and one of the authors\cite{ais}.}

We introduced a recipe whereby we could relate a topological link to an appropriate   entangled quantum state, via a unitary representation of the braid word producing the link.  For the two cases of links produced by $B_3$, the Borromean rings link and the one we dubbed the NUS link, we used two different unitary representations of $B_3$.  It should come as no surprise that different unitary representations produce different pictures of entanglement, as quantum entanglement is not invariant under unitary transformations.
And indeed, from our description of the Borromean rings link and the NUS link in the Introduction, we can see that the topological entanglement properties of these two links are quite different.  Similarly, from the discussion following Eqs.(\ref{GHZ}) and (\ref{phi}) we also see that the quantum entanglement properties of the states $|GHZ\rangle$ and $|\phi\rangle$ are similarly distinct.

Further, it would appear that the entanglement properties in the 3-qubit case are not invariant under {\em local unitary transformations} either. It  has been pointed out\cite{ZJG} that in fact the two states $|GHZ\rangle$ and $|\phi\rangle$ are locally unitarily equivalent, since
by use of the {\em local transformation}
$V = v \otimes v \otimes v$ where $v= \frac{1}{\sqrt{2}}\left(
                                        \begin{array}{cc}
                                          1 & 1 \\
                                          -1 & 1 \\
                                        \end{array}
                                      \right)$,
$$V \frac{1}{\sqrt{2}}(|0,0,0\rangle+|1,1,1\rangle) = -\frac{1}{2}(|0,0,0\rangle+|0,1,1\rangle+|1,0,1\rangle+|1,1,0\rangle).$$
Thus, in the case of {\em tripartite} states, at least, local unitary equivalence does not preserve  the entanglement properties.

\section*{Acknowledgments}
Both authors wish to acknowledge  discussions with  Professor Choo
Hiap Oh of the  Department of Physics  whom we thank for his warm
hospitality, as well as that  of the Centre for Quantum
Technologies at the National University of Singapore. This work is
supported in part by Singapore's A*STAR grant WBS (Project Account
No.) R-144-000-189-305, and in part by the National Science
Council (NSC) of the R.O.C. under Grant No. NSC
96-2112-M-032-007-MY3.

\end{document}